\documentclass[letterpaper, 10 pt, conference]{ieeeconf}  

\IEEEoverridecommandlockouts                              
\usepackage{graphicx,scalefnt}
\usepackage{amsmath,graphicx,scalefnt}
\usepackage{graphicx}
\usepackage{caption}
\usepackage{subcaption}
\usepackage{float}
\usepackage{subfig}
\title{\LARGE \bf Filter Bank Common Spatial Patterns in Mental Workload Estimation}

\author{Mahnaz Arvaneh$^{1}$, Alberto Umilta $^{2}$, and Ian H. Robertson$^{1}$ 
\thanks{*This work was supported by Science Foundation Ireland (SFI), under
Grant No. 12/RC/2289.}
\thanks{$^{1}$ M. Arvaneh and I. H. Robertson are with Trinity College Institute of Neuroscience and Insight Centre for Data Analytics,  Dublin, Ireland,
        (emails:arvanehm,iroberts@tcd.ie)}%
\thanks{$^{2}$ A. Umilta is
with School of Psychology, University of Padova, Padova, Italy,
(email:alberto.umilta@gmail.com).}%
}

\begin{document}

\maketitle
\thispagestyle{empty}
\pagestyle{empty}

\begin{abstract}

EEG-based workload estimation technology provides a real time means
of assessing mental workload. Such technology can effectively
enhance the performance of the human-machine interaction and the
learning process. When designing workload estimation algorithms, a
crucial signal processing component is the feature extraction step.
Despite several studies on this field, the spatial properties of the
EEG signals were mostly neglected. Since EEG inherently has a poor
spacial resolution, features extracted individually from each EEG
channel may not be sufficiently efficient. This problem becomes more
pronounced when we use low-cost but convenient EEG sensors with
limited stability which is the case in practical scenarios. To
address this issue, in this paper, we introduce a filter bank common
spatial patterns algorithm combined with a feature selection method
to extract spatio-spectral features discriminating different mental
workload levels. To evaluate the proposed algorithm, we carry out a
comparative analysis between two representative types of working
memory tasks using data recorded from an Emotiv EPOC headset which
is a mobile low-cost EEG recording device. The experimental results
showed that the proposed spatial filtering algorithm outperformed
the state-of-the algorithms in terms of the classification accuracy.

\end{abstract}

\section{INTRODUCTION}


Recent advances in sensor technologies and computational algorithms
make it possible to non-invasively monitor brain activities and
mental states. In particular, real-time assessment of mental
workload (MW) has attracted a lot of attentions. Assessing MW can be
beneficial in applications requiring high level of engagement,
concentration and alertness such as aviation, driving, education and
industrial production lines \cite{cit:1,cit:2}.

In complex high demanding tasks, the human's performance might drop
due to the mental overload caused by excessive amount of information
to be processed. In contrast, human tends to make errors when MW is
kept in a lower level than the proper level due to getting bored.
Thus, to achieve the best performance, the flow of the information
and the complexity of the task should be controlled by correctly
estimating the user's MW \cite{cit:1,cit:2,cit:3}. In addition, a
system that provides real-time feedback based on the detected MW
might potentially enhance the cognitive performance and the learning
process by encouraging the user to stay focused and engaged
\cite{cit:4}.

It is not still clear how to exactly define MW \cite{cit:5}.
However, it is well accepted that MW is correlated with task demand,
time pressure, person's capacity and his/her performance
\cite{cit:6}. Thus, in the existing studies, generally well-defined
cognitive tasks were used in different difficulty and demand levels
to manipulate a person's MW level. Simultaneously, a range of
different physiological signals were recorded for estimating the
subject's MW, such as pupil size, eye blink, skin conductance,
electrocardiogram (ECG), and electroencephalogram (EEG).
Interestingly, extensive comparisons reported by different research
groups revealed that EEG is the most promising signal for estimating
MW \cite{cit:7}.

Studies that used EEG mostly achieved satisfactory results based on
band power features \cite{cit:8,cit:9}. It is shown that theta
(4-8Hz) and alpha (8-12Hz) are particularly sensitive to changes in
MW \cite{cit:8,cit:9,cit:10}. Typically, theta in the frontal
midline regions of the scalp increases as task demands increase
\cite{cit:10}, while alpha decreases in parietal regions when the
workload increases \cite{cit:8,cit:9}. In addition to theta and
alpha, MW can also influence other frequency bands (e.g. gamma and
beta) in some subjects. Indeed, the exact locations and the
frequency bands affected by MW vary between subjects and tasks
\cite{cit:11,cit:12}.

Despite several studies on the EEG-based estimation of MW, the
spatial properties of the EEG signals were mostly neglected, and
features were extracted from each channel individually. However, due
to volume conduction, EEG has inherently a poor spatial resolution.
Hence, applying proper spatial filters increases the signal to noise
ratio, and possibly leads to a more accurate MW estimation.
Importantly, the previous studies mostly relied on costly wired EEG
equipments which require injecting gel on the scalp to have good
quality signals. In order to use MW estimator tools in our daily
life, they need to be utilized with dry/non-gel wireless EEG sensors
with limited stability. Thus, the impact of applying proper spatial
filtering algorithms could be even more pronounced in practical
scenarios.

To address the aforementioned issue, we first introduce a filter
bank common spatial patterns (FBCSP) technique \cite{cit:13}
combined with a feature selection method to extract EEG
spatio-spectral features discriminating MW levels. Second, we look
into practical issues by conducting a comparative study between two
working memory tasks (i.e. two different n-back tasks) with three MW
levels. The brain signals were recorded from 6 subjects using a
wireless low-cost Emotiv EPOC headset \cite{cit:13_1}. We compare
the performance of the proposed FBCSP algorithm in classification of
different MW levels with classification results obtained from EEG
band power features. The between session classification accuracies
are calculated based on different time intervals, while the training
time is kept less than 6 minutes per MW level.

\section{MATERIALS AND METHODS}
\subsection{Experimental design}
In total, 6 young adults aged 19-33 years were participated in this
study. All the participants gave their informed consent to the study
which had been reviewed and approved by the ethical review board of
the School of Psychology, Trinity College Dublin, in accordance with
the Declaration of Helsinki. The participants were asked to complete
4 sessions of a verbal n-back task and 4 sessions of a spatial
n-back task. Each session consisted of three 2-min blocks (i.e.
0-back, 1-back and 2-back). After each block, the participants took
a 12-15 seconds rest. They were also welcome to take a break at the
end of each session, if they desired. Each block consisted of 60
trials, where each trial started by 500 ms presentation of the
stimulus followed by 1500 ms inter stimulus interval.

\begin{figure}[htb]
\begin{minipage}[b]{1\linewidth}
  \centering
  \centerline{\includegraphics[width=6.5cm,height=1.5cm]{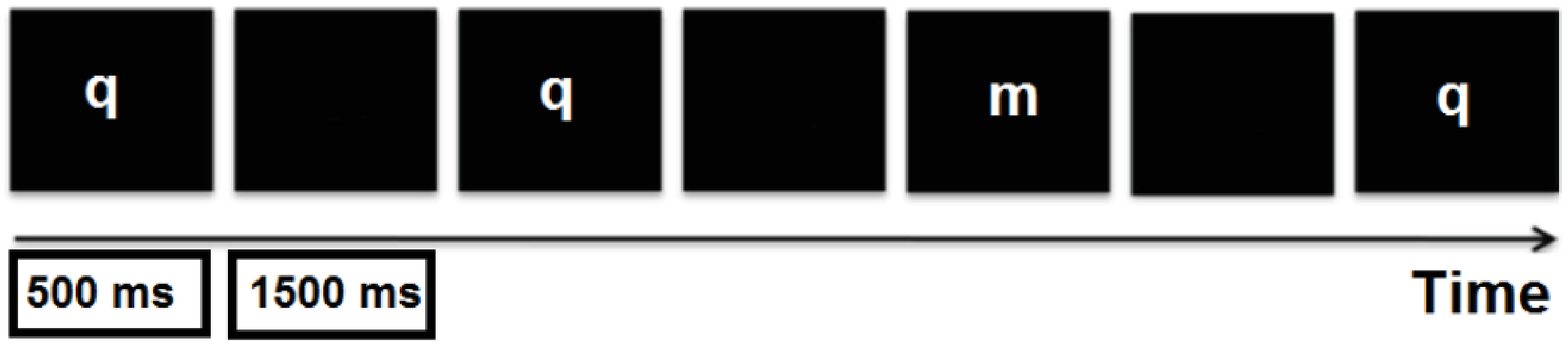}}
 \vspace{-0.2cm}
 \centerline{\small{(a) Verbal n-back task}}\medskip
\end{minipage}
\hfill
\\
\begin{minipage}[b]{1\linewidth}
  \centering
  \centerline{\includegraphics[width=6.5cm,height=1.5cm]{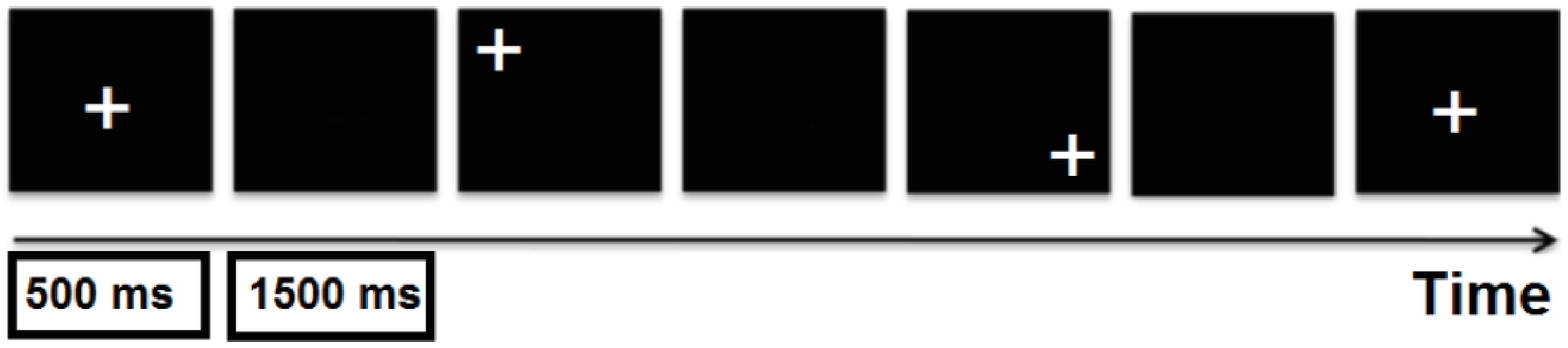}}
\vspace{-0.2cm}
  \centerline{\small{(b) Spatial n-back task}}\medskip
\end{minipage}
\vspace{-.5cm}
 \caption{\small{Graphical representation of the n-back tasks used in this study}}
\label{fig:task}
\end{figure}

The verbal and the spatial n-back tasks performed in this study were
similar to the tasks introduced in \cite{cit:9}. In each block of
the verbal n-back task, a series of letters was randomly presented
at the center of the screen (see Fig. \ref{fig:task}.a). The
participants were asked to remember the new letter and respond if it
was the same as the letter presented n trials before. In total, 5
consonant English letters were used in the verbal task. In the
spatial n-back task, a white cross was presented randomly in 5
different locations on the screen (see Fig. \ref{fig:task}.b).
Participants were asked to compare the current location of the cross
to that occurred n trials before, and respond if they were same. In
the 0-back blocks, the participants were only required to respond to
those stimuli (i.e. letters/locations) that were the same as the one
presented at the beginning of the block. In all the blocks, 20\% of
the stimuli were targets. To reduce the learning effects, before
starting the test, the participants practiced until they reached the
satisfactory performance.

\subsection{EEG data acquisition}
EEG was acquired using an Emotiv EPOC headset \cite{cit:13_1}. The
Emotiv EPOC headset is completely wireless with 14 electrodes and 2
mastoid reference electrodes. In this study, we used 12 of the 14
available electrodes, namely F3, F4, F7, F8, FC5, FC6, T7, T8, P7,
P8, O1, O2, as well as the two mastoid electrodes. Saline liquid was
used to reduce the impedance of the electrodes to a satisfactory
level. The sampling rate was 128 Hz. The recorded EEG data were
segmented to the intervals of 2, 4, and 6 s starting from the onset
of the stimuli. The segments with amplitudes exceeding $+75\mu V$,
or voltage steps of more than 150$\mu V$ within a window of 200 ms
were rejected from further analysis.

\subsection{Proposed EEG-based mental workload estimator} \label{sec:FBCSFS}
FBCSP is extensively used in classification of EEG-based motor
imagery data \cite{cit:13}. In this study, the FBCSP algorithm was
used to extract spatio-spectral features discriminating the MW
levels. Thereafter, a feature selection method was applied to select
the most discriminative set of features. Finally, a naive bayesian
classifier was used for classification. The details about the
proposed MW estimation algorithm are as follows:


\vspace{0.1cm} 1) Multi-band spectral filtering: A filter bank was
applied to decompose the EEG data into nine equal frequency bands,
namely 4-8, 8-12, ..., 36-40 Hz. These frequency ranges cover all
the commonly used frequency bands in the classification of MW.

\vspace{0.1cm} 2) Common spatial patterns (CSP): The EEG data from
each frequency band were spatially filtered using the CSP filters
\cite{cit:13}. Among various spatial filters, CSP has been highly
successful in classification of two classes of EEG data
\cite{cit:14}. CSP increases the discrimination between two classes
by maximizing the variance of one class while the variance of the
other class is minimized.

Let $\mathbf{X}\!\in\!\mathbf{R}^{N_{c} \times S}$ denote a bandpass
filtered single-trial EEG data, where $N_{c}$ and $S$ are the number
of channels and the number of measurement samples respectively. The
CSP transformation matrix, $\mathbf{W}\!\in\!\mathbf{R}^{N_{c}
\times N_{c}}$, linearly transforms $\mathbf{X}$ as
$\mathbf{Z}=\mathbf{W}\mathbf{X}$. $\mathbf{W}$ is generally
computed by solving the eigenvalue decomposition problem: 
\begin{equation} \label{eqn:CSP2}
\mathbf{C}_{1}\mathbf{W} = \left(\mathbf{C}_{1} + \mathbf{C}_{2}
\right)\mathbf{W}\mathbf{D},
\end{equation}
where $\mathbf{C}_{1}$ and $\mathbf{C}_{2}$ are respectively the
averaged covariance matrices of the bandpass filtered EEG data
obtained from each class; $\mathbf{D}$ is the diagonal matrix that
contains the eigenvalues of
$(\mathbf{C}_{1}+\mathbf{C}_{2})^{-1}\mathbf{C}_{1}$. Usually, only
the first and the last $m$ rows of $\mathbf{W}$ are used as the most
discriminative filters to perform spatial filtering \cite{cit:15}.

\vspace{0.1cm} 3) Feature extraction: The spatio-spectrally filtered
EEG data were used to determine the features associated to each
band-pass frequency range. Based on the Ramoser formula
\cite{cit:15}, the features of the $k^{\mathrm{th}}$ trial of the
EEG data belonging to each frequency band were calculated as
\begin{equation}\label{eq2}
\mathbf{v}_{k}=\mathrm{log}(\mathrm{diag}(\mathbf{Z}_{k}\mathbf{Z}^{\mathrm{T}}_{k})/\mathrm{trace}[\mathbf{Z}_{k}\mathbf{Z}^{\mathrm{T}}_{k}]),
\end{equation}
where $\mathbf{v}_{k}\!\in\!\mathbf{R}^{1\times 2m}$; diag(.)
returns the diagonal elements of the square matrix; and the
superscript $\mathrm{T}$ denotes the transpose of the matrix. Since
we have nine frequency bands, the total number of features for each
trial was $9\times2m$. In this study, m was set to two.

\vspace{0.1cm} 4) Feature selection: The mutual information
algorithm was used for ranking the features. Subsequently, the top
$n$ ranked features were used for classification. The value $n$ was
chosen based on 10-fold cross-validation on the training data, such
that the top $n$ features yielding the highest average
cross-validation accuracy on the training data were selected as the
most discriminative set of features for the MV classification.


\section{Result}
\subsection{Behavioral performance}
For the verbal and the spatial n-back tasks, we performed 3
(Difficulty: 0-back vs. 1-back vs. 2-back) $\times$ 2 (Task: verbal
vs. spatial) repeated ANOVA tests on both the error rates and the
response times. We observed significant main effects of Difficulty
on both the error rate ($F(2,15)=8.757, p=0.006$) and the response
time ($F(2,15)=21.75, p<0.001$). The response time increased by
increasing the memory load with averages of 0.432s, 0.515s, 0.661s
respectively for the verbal n-back task, and 0.433s, 0.515s, and
0.702s respectively for the spatial n-back task. Similarly, the
error rate increased by increasing the difficulty with averages of
0.83\%,1.5\%,5.5\% respectively for the verbal n-back task, and
0.17\%, 1.17\% and 4.83\% respectively for the spatial n-back task.
Neither Task nor the interaction between Task and Difficulty was
significant. Post-hoc tests showed that the error rates and the
response time were significantly different in all the difficulty
levels. These results suggest that the tasks successfully induced
three different MW levels in the participants.

\subsection{Effects of spacial filtering and feature selection on
MW estimation}

To consider the effects of the proposed algorithm on the MW
estimation, four different classification models were trained. The
first model (abbreviated as FBCSP(FS)) was obtained based on the
algorithm described in Section (\ref{sec:FBCSFS}). Indeed, in this
model, FBCSP was used to extract spatio-spectral features.
Thereafter, the best set of features was selected using the proposed
feature selection method. In the second model (abbreviated as
FBCSP(AllF)), all the features obtained from FBCSP were used for
classification (i.e. without any feature selection). In the third
model (abbreviated as BP(AllF)), the band power features were
obtained per each channel using the 9 frequency bands (i.e. 4-8,
8-12,..., 36-40 Hz) without applying any spacial filtering or
feature selection algorithms. In fact, these 9 frequency bands are
the same as those employed in FBCSP. BP(AllF) is similar to the
models that are commonly used in MW estimation studies
\cite{cit:8,cit:9}. In the last model (abbreviated as BP(FS)), using
the proposed feature selection method, a subset of the band power
features obtained from the third model was used for classification.
For each n-back task, the first three sessions were used for
training the classification models and the last session was used for
evaluation. The results presented in this subsection are based on 2
seconds EEG intervals extracted from the onset of the stimuli.

\begin{figure}[htb]
\begin{minipage}[b]{0.495\linewidth}
  \centering
  \centerline{\includegraphics[width=4.8cm,height=3.5cm]{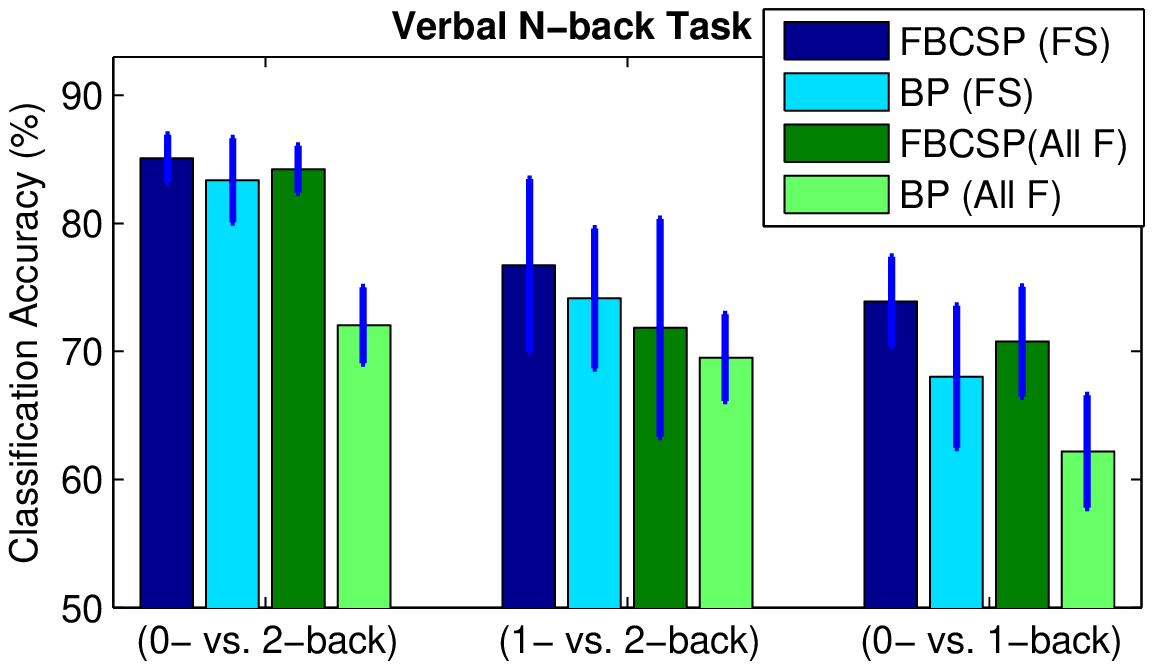}}
 \centerline{(a)}\medskip
\end{minipage}
\hfill
\begin{minipage}[b]{0.495\linewidth}
  \centering
  \centerline{\includegraphics[width=4.8cm,height=3.5cm]{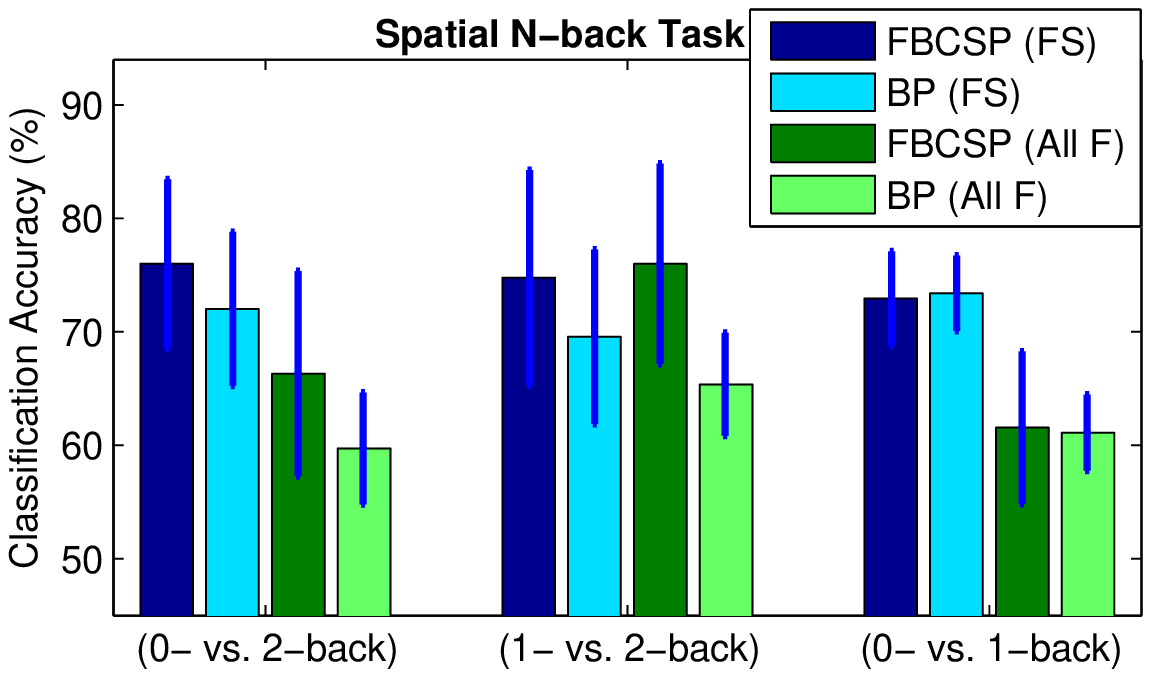}}
  \centerline{(b)}\medskip
\end{minipage}
\vspace{-.5cm}
 \caption{\small{Average classification accuracies of (a) verbal and (b) spatial n-back tasks obtained using 4 different models.
 The window size is 2 sec. BP, All F and SF denote the band power features, all the features, and the selected features
 respectively.}}
\label{fig:bar}
\end{figure}

Fig. \ref{fig:bar} shows the classification results of the four
models under the three conditions for the verbal and the spatial
n-back tasks. All the four models achieved the classification
accuracies above the chance level. This confirms the satisfactory
quality of the EEG signals recorded by the Emotiv EPOC headset. As
shown in Fig. \ref{fig:bar}, on average FBCSP(FS) outperformed all
the other models, whereas the BP(AllF) performed the worst.
Performing 4 (Models) $\times$ 3 (Difficulty: 0-back vs. 1-back vs.
2-back) repeated ANOVA tests revealed significant main effects of
the models in the letter ($F(3,15)=11.35, P=0.001$) and the spatial
($F(3,15)=4.86, p=0.04$) n-back tasks, respectively. A close to
significant main effect of Difficulty was also observed in the
letter n-back task ($F(2,10)=4.06, p=0.051$). Importantly, Post-hoc
tests showed that the proposed FBCSP(FS) algorithm significantly
performed better than the BP(AllF) algorithm which is commonly used
in MW estimation.

Fig. \ref{fig:bar}.a shows that in the verbal task the low MW (i.e.
0-back) was separated from the high MW (2-back) with the highest
average accuracy, while the classification between the low MW (i.e.
0-back) vs. the medium MW was (i.e. 1-back) the least accurate among
the other conditions. Paired t-tests showed that the classification
results of the proposed FBCSP(FS) were significantly different
between the (0- vs. 2-back) and (0- vs. 1-back) conditions (p=0.03).
Unlike the verbal n-back task, in the spatial n-back task the
classification accuracies of the proposed FBCSP(FS) are closer over
the three conditions, although still the highest accuracy obtained
in the classification of 0- vs. 2-back.

\subsection{Effects of EEG window size}

The results obtained in the  previous subsection were all based on 2
seconds EEG windows. To consider the effects of EEG window size on
the MW classification accuracy, the performance of the proposed
FBCSP(FS) algorithm was also evaluated using longer EEG intervals
(i.e. 4 and 6 seconds).

\begin{figure}[htb]
\begin{minipage}[b]{0.49\linewidth}
  \centering
  \centerline{\includegraphics[width=4.3cm,height=3.7cm]{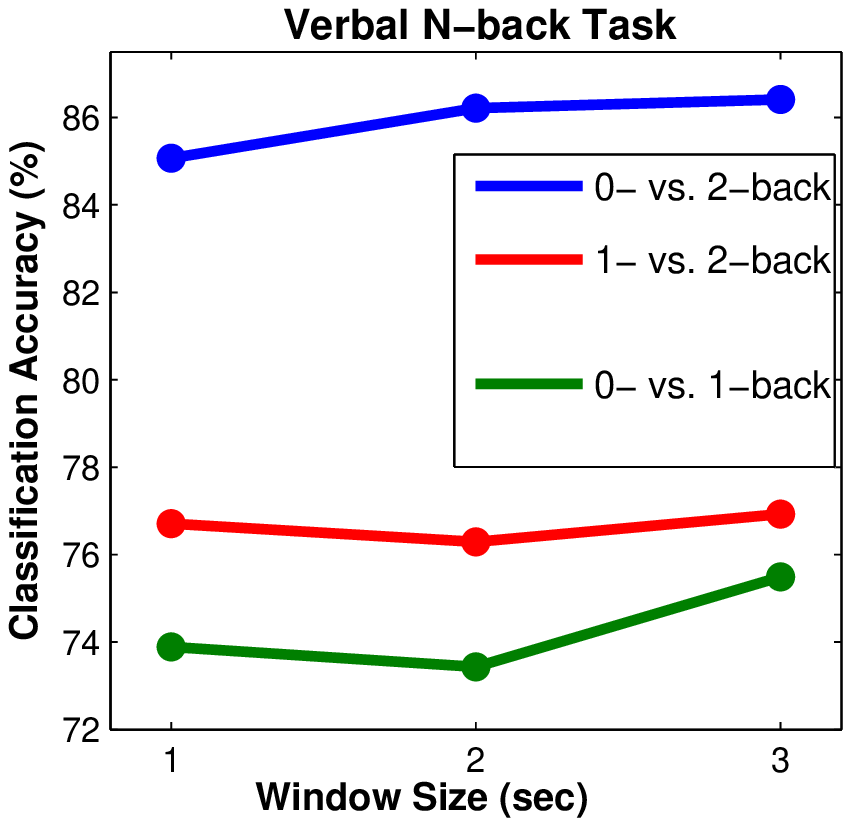}}
 \centerline{}\medskip
\end{minipage}
\hfill
\begin{minipage}[b]{0.49\linewidth}
  \centering
  \centerline{\includegraphics[width=4.3cm,height=3.7cm]{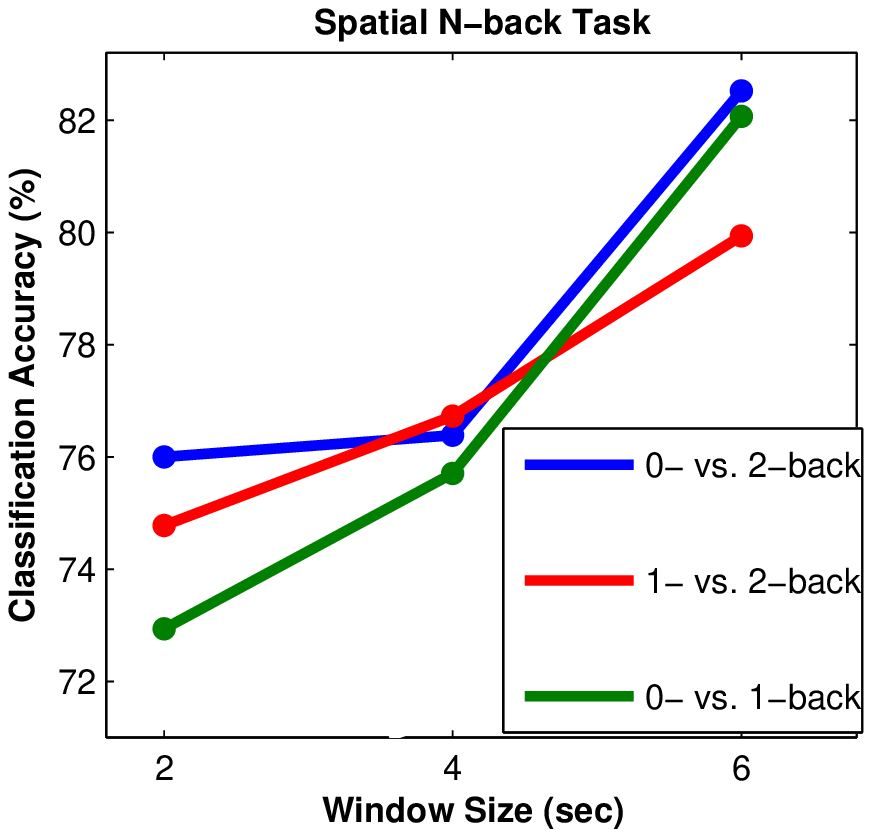}}
  \centerline{}\medskip
\end{minipage}
\vspace{-0.8cm}
 \caption{\small{Average classification accuracy of the proposed FBCSP(FS) algorithm as a function of window size,
 for (a) the verbal and (b) the spatial n-back tasks.}}
\label{fig:Win_size}
\end{figure}

As shown in Fig. \ref{fig:Win_size}, the classification accuracy
improved when the EEG window size increased. Interestingly, this
improvement was more pronounced in the spatial n-back task. A
repeated ANOVA test revealed a significant main effect of the window
size on the accuracy ($F(2,10)=11.24,p=0.003$) in the spatial n-back
task. However, in the verbal n-back task, the window size did not
have a significant effect on the accuracy ($F(2,10)=0.34,p=0.72$).
It should be noted that longer window size means higher chance of
having blinks or muscle artifacts. Thus, due to artifact rejection,
increasing the window size leads to a smaller number of trials left
for training. This might negatively affect the results as the
accurate estimation of the CSP matrix is associated with the
training size \cite{cit:15}. Considering this issue, in future it
should be further investigated why increasing the window size did
not bring a large advantage for the verbal n-back task.

\subsection{Spatial filters in different frequency bands}

\begin{figure}[htb]
\begin{minipage}[b]{0.49\linewidth}
  \centering
  \centerline{\includegraphics[width=4.3cm,height=3.3cm]{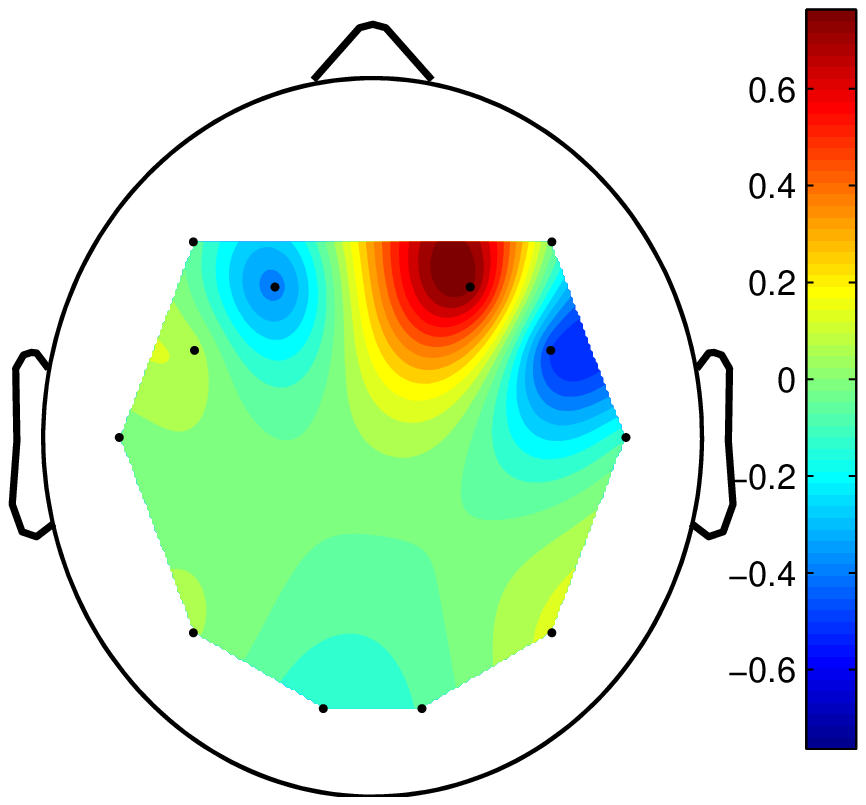}}
 \centerline{(a)}\medskip
\end{minipage}
\hfill
\begin{minipage}[b]{0.49\linewidth}
  \centering
  \centerline{\includegraphics[width=4.3cm,height=3.3cm]{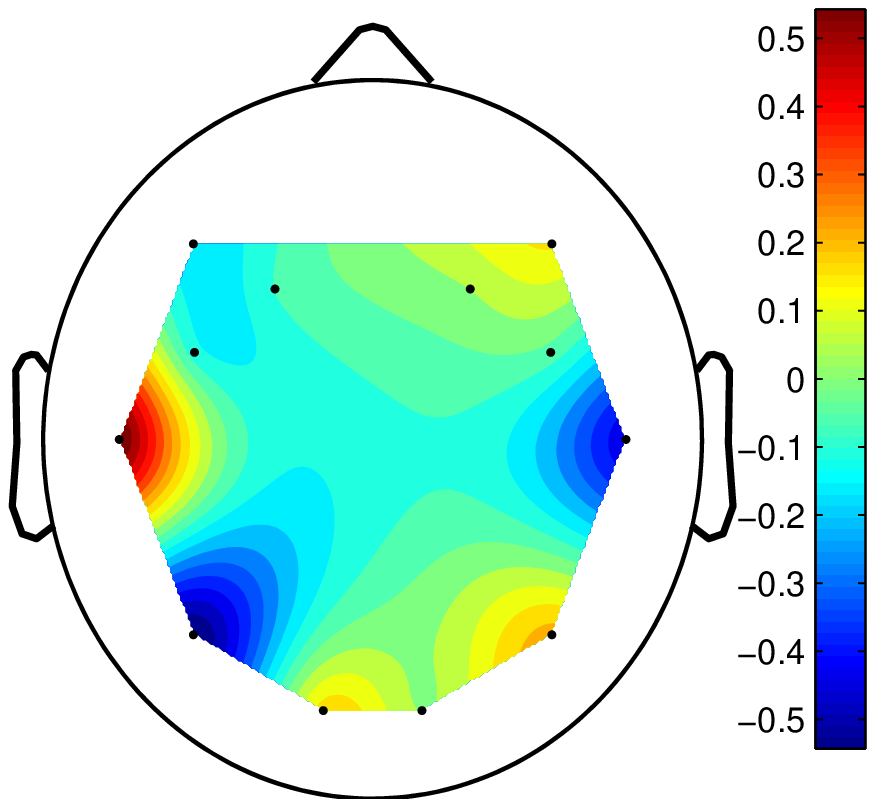}}
  \centerline{(b)}\medskip
\end{minipage}
\vspace{-0.4cm}
 \caption{\small{Spatial filters obtained for (a) theta and (b) alpha frequency bands in the verbal n-back task, for one subject.}} \label{fig:spat_filter}
\end{figure}

To better understand why the proposed algorithm improved the
classification results, two spatial filters obtained for one of the
subjects were presented in Fig. \ref{fig:spat_filter}. The spatial
filters were trained in order to get an optimum discrimination
between 0- and 2-back conditions in the letter task. As shown in
Fig. \ref{fig:spat_filter}, the spatial filter obtained for the
theta rhythm (4-8 Hz) gives more weights to the frontal electrodes,
while attenuates the effects of the other channels.  In the same
line, the spatial filter obtained for the alpha rhythm (8-12 Hz) is
more focused on the temporal and parietal electrodes, while the
effects of the other channels are mitigated. Thus, by adding spatial
filters to the MW classification algorithms, the effects of
irrelevant and redundant channels that might be different from band
to band are attenuated, and more neurophysiologically relevant
features are extracted.

\section{CONCLUSIONS}

To create a classification model that accurately estimates mental
workload in practical scenarios, the reliability of the system
should be evaluated using convenient low-cost EEG sensors with
limited stability. In such a noisy environment, using spatial
filters could be crucial in improving signal to noise ratio. To
address these issues, we introduced a filter bank common spatial
patterns algorithm combined with a feature selection method to
extract spatio-spectral features discriminating different mental
workloads. We compared 2 representative working memory tasks: the
verbal and the spatial n-back tasks using data collected from Emotiv
EPOC, a widely used wireless EEG headset. Our experimental results
showed that the proposed spatio-spectral features outperformed the
state-of-the art algorithms in classification of different workload
conditions in both tasks. The results also showed that spatial
filters could improve the accuracy of the MW classification
algorithms by attenuating the effects of irrelevant and redundant
channels, and enhancing the influence of the neurophysiologically
relevant channels.


\begin{thebibliography}{99}

\bibitem{cit:1}
T. O. Zander, C. Kothe, S. Jatzev, and M. Gaertner, "Enhancing
human-computer interaction with input from active and passive
brain-computer interfaces," \emph{Brain-Computer Interfaces},
Springer London, 2010, pp. 181-199.

\bibitem{cit:2}
D. B. Kaber, E. Onal, and M. R. Endsley, "Design of automation for
telerobots and the effect on performance, operator situation
awareness, and subjective workload," \emph{Human Factors and
Ergonomics in Manufacturing}, vol. 10, no. 4, pp. 409–30, 2000.

\bibitem{cit:3}
G. F. Wilson, and C. A. Russell, "Operator functional state
classification using multiple psychophysiological features in an air
traffic control task," \emph{Human Factors: The Journal of the Human
Factors and Ergonomics Society}, vol. 45, no. 3, pp. 381-389, 2003.

\bibitem{cit:4}
A. Holm, K. Lukander, J. Korpela, M. Sallinen, and K. M. I.
M\"{u}ller, "Estimating brain load from the EEG," \emph{The
Scientific World Journal}, vol. 9, pp. 639-651, 2009.

\bibitem{cit:5}
De Jong, T. "Cognitive load theory, educational research, and
instructional design: some food for thought," \emph{Instructional
Science}, vol. 38, no. 2, 105-134, 2010.

\bibitem{cit:6}
F. G. Paas, J. J. Van Merri\"{e}nboer, and J. J. Adam, "Measurement
of cognitive load in instructional research," \emph{Perceptual and
motor skills}, vol. 79, no. 1, pp. 419-430, 1994.

\bibitem{cit:7}
 M. A. Hogervorst, A. M. Brouwer, and J. B. van Erp, "Combining and
comparing EEG, peripheral physiology and eye-related measures for
the assessment of mental workload," \emph{Frontiers in
neuroscience}, vol. 8, 2014.

\bibitem{cit:8}
A. M. Brouwer et al., "Estimating workload using EEG spectral power
and ERPs in the n-back task," \emph{Journal of Neural Engineering},
vol. 9, no. 4, 2010.

\bibitem{cit:9}
D. Grimes et al, "Feasibility and pragmatics of classifying working
memory load with an electroencephalograph," in \emph{SIGCHI
Conference on Human Factors in Computing Systems}, April 2008, pp.
835-844.

\bibitem{cit:10}
J. Onton, A. Delorme, and S. Makeig, "Frontal midline dynamics
during working memory," \emph{Neuroimage}, vol. 27, pp. 341-356,
2005.

\bibitem{cit:11}
J. B. Brookings, G. F. Wilson, and C. R. Swain, "Psychophysiological
responses to changes in workload during simulated air traffic
control," \emph{Biol. Psychol}. vol. 42, pp.361–77, 1996.

\bibitem{cit:12}
M. W. Howard et al., "Gamma oscillations correlate with working
memory load in humans," \emph{Cerebral Cortex}, vol. 13, pp.
1369-74, 2003.

\bibitem{cit:13}
K. K. Ang, Z. Y. Chin, H. Zhang, and C. Guan, "Mutual
information-based selection of optimal spatial-temporal patterns for
single-trial EEG-based BCIs," \emph{Pattern Recognition}, vol. 45,
no. 6, pp. 2137-44, 2012.

\bibitem{cit:13_1}
EmotivSystems. Emotiv - brain computer interface technology.
http://emotiv.com.

\bibitem{cit:14}
B. Blankertz, R. Tomioka, S. Lemm, M. Kawanabe, and K. R.
M\"{u}ller, "Optimizing spatial filters for robust EEG single-trial
analysis," \emph{IEEE Signal Process. Mag.}, vol. 25, no. 1, pp.
41-56, 2008.

\bibitem{cit:15}
H. Ramoser, J. M\"{u}ller-Gerking, and G. Pfurtscheller, "Optimal
spatial filtering of single trial EEG during imagined hand
movement," \emph{IEEE Trans. Rehabil. Eng.}, vol. 8, no. 4, pp.
441-446, Dec. 2000.


\end{thebibliography}
\end{document}